\documentclass[runningheads]{llncs}
\usepackage{graphicx}

\usepackage{adjustbox}
\usepackage{multirow}
\usepackage{amsmath}
\usepackage{comment}
\begin{document}
\title{Segmenting Fetal Head with Efficient Fine-tuning Strategies in Low-resource Settings: an empirical study with U-Net}
%
%
\author{Fangyijie Wang\inst{1,2}\orcidID{0009-0003-0427-368X} \and
Gu\'enol\'e Silvestre\inst{1,3} \and
Kathleen M. Curran\inst{1,2}\orcidID{0000-0003-0095-9337}}
\authorrunning{Author et al.}
%
\institute{Science Foundation Ireland Centre for Research Training in Machine Learning 
\email{fangyijie.wang@ucdconnect.ie}\\ \and
School of Medicine, University College Dublin, Dublin, Ireland \and
School of Computer Science, University College Dublin, Dublin, Ireland
}
\maketitle              
\begin{abstract}

Accurate measurement of fetal head circumference is crucial for estimating fetal growth during routine prenatal screening. Prior to measurement, it is necessary to accurately identify and segment the region of interest, specifically the fetal head, in ultrasound images. Recent advancements in deep learning techniques have shown significant progress in segmenting the fetal head using encoder-decoder models. Among these models, U-Net has become a standard approach for accurate segmentation. However, training an encoder-decoder model can be a time-consuming process that demands substantial computational resources. Moreover, fine-tuning these models is particularly challenging when there is a limited amount of data available. There are still no “best-practice” guidelines for optimal fine-tuning of U-net for fetal ultrasound image segmentation. This work summarizes existing fine-tuning strategies with various backbone architectures, model components, and fine-tuning strategies across ultrasound data from Netherlands, Spain, Malawi, Egypt and Algeria. Our study shows that (1) fine-tuning U-Net leads to better performance than training from scratch, (2) fine-tuning strategies in decoder are superior to other strategies, (3) network architecture with less number of parameters can achieve similar or better performance. We also demonstrate the effectiveness of fine-tuning strategies in low-resource settings and further expand our experiments into few-shot learning. Lastly, we publicly released our code and specific fine-tuned weights.


\end{abstract}
\section{Introduction}

Ultrasound (US) imaging is widely used for the diagnosis, screening and treatment of a large number of diseases because of its portability, low cost and non-invasive nature \cite{Zaffino:2020}. In recent decades, US screening has become a common method used for prenatal evaluation of fetal growth, fetal anatomy, and estimation of gestational age, as well as monitoring pregnancy \cite{Loughna:2009,Salomon:2011}. The Head Circumference (HC) is a fundamental measurement used to estimate fetal size \cite{Salomon:2011}. When expert sonographers perform fetal HC measurements manually, an ellipse around the perimeter of the fetal skull is defined as the region of interest (ROI) \cite{poojari:2022}. The circumference of this ROI is subsequently calculated as the fetal head circumference. However, this process is patient-specific, operator-dependent, machine specific, and prone to intra and inter-user variability, which can result in inaccurate measurements \cite{Sarris:2012}. Therefore, the calculation of fetal size by head circumference measurement becomes one of the most challenging fetal biometry to work with. Regular training and audits are highly recommended for professional development and to maintain competency \cite{Milner:2018}.

Recently, deep learning (DL) segmentation techniques have been developed to achieve high precision outcomes in semantic segmentation tasks. Ronneberger et al. \cite{Ronneberger:2015} propose the U-Net architecture for improving the efficiency of biomedical image segmentation tasks with annotated samples. However, we observe that training a U-Net, particularly from scratch, poses a significant challenge due to scarcity of labelled data. 
Xie et al. \cite{Xie:2020} implement a Computer-Aided Diagnosis (CAD) system using Transfer Learning (TL) techniques using a significant amount of fetal brain images from US videos. While annotated data were available for this study, a substantial amount of labeled data is typically not readily available, especially in low-resource settings (LRS). LRS are characterized by a lack of adequate healthcare resources and systems that fail to meet recognized global standards, while high-resource settings (HRS) present the opposite environment \cite{Piaggio:2021}.
Cheng et al. \cite{cheng:2021} utilize cross-domain TL with U-Net architecture for precise and fast image segmentation. This method requires a U-Net with an unfrozen bottleneck layer pre-trained on non-ultrasound data set XPIE \cite{Xia:2017}. 

This paper presents a semantic segmentation technique for fetal head US images using deep neural networks, employing several fine-tuning strategies. To perform semantic segmentation on fetal US images, we fine tune a U-Net \cite{Ronneberger:2015} network, with a pre-trained MobileNet \cite{Howard:2018} as its encoder. MobileNet is a lightweight pre-trained CNN model, which has the ability to develop an efficient method using fine-tuning techniques. We investigate the effect of fine-tuning on various decoder layers for segmentation of the fetal head. To assess the generalization and transferability of different fine-tuning strategies, we train the model on fetal US data from HRS to previously unseen fetal US data acquired in LRS. 

The main contributions of this paper are: (i) An effective fine-tuning strategy for segmenting the fetal head from US data that achieves competitive accuracy in LRS with minimal computing resources. (ii) We demonstrate the effectiveness of the fine-tuning strategy proposed as a zero-shot learning approach for fetal head segmentation across three African countries. (iii) This is the first study to perform generalization and transferable fine tuning strategies in high and low resource settings.

\section{Methods}

\begin{figure}[!htb]
    \centering
    \includegraphics[width=\textwidth]{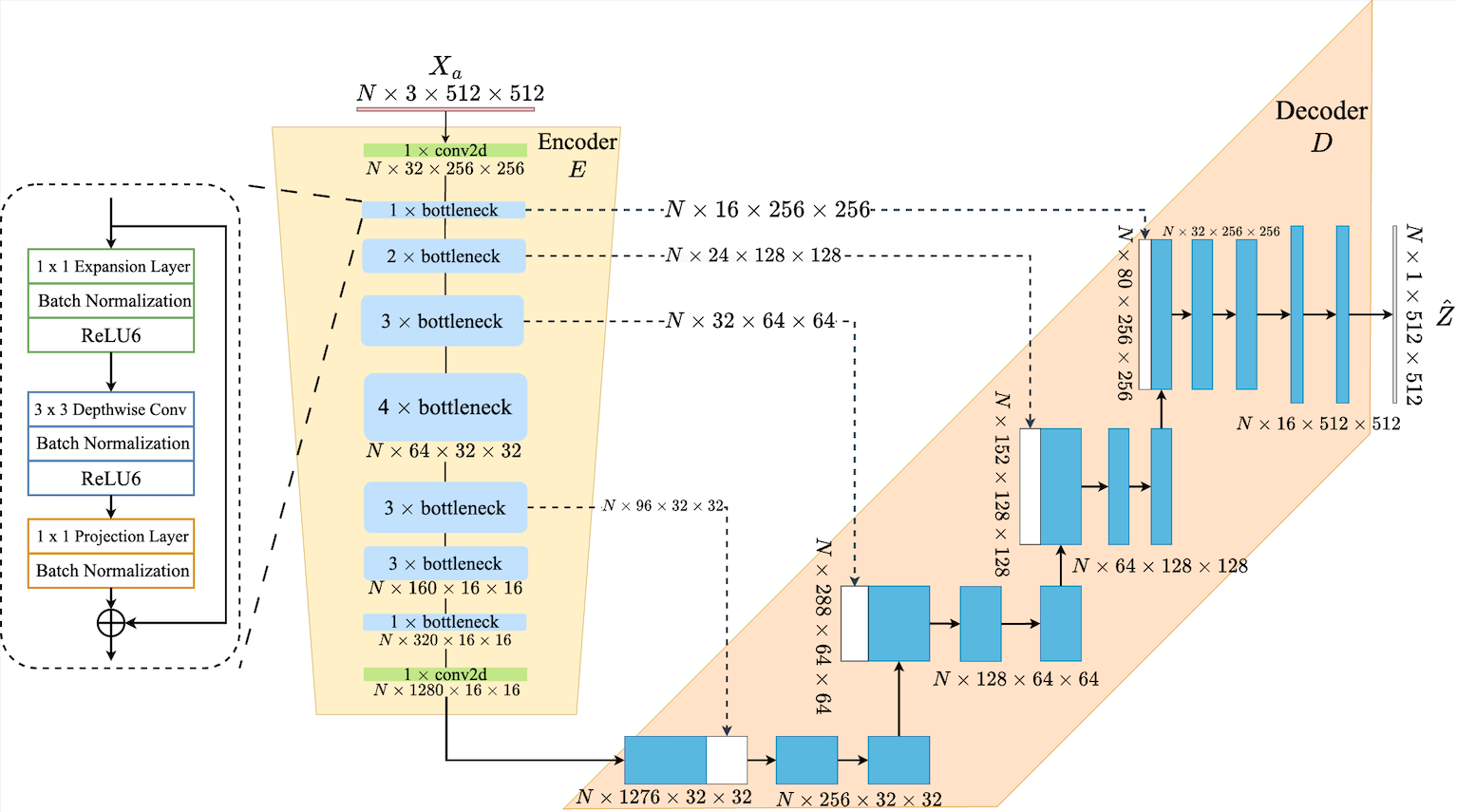}
    \caption{Overview of the U-Net network with encoder ($E$) and decoder ($D$). The encoder stack is replaced with MobileNet v2. The features from layers within $E$ are concatenated to features in $D$ by skip connections. The \textbf{left} figure illustrates the details of the bottleneck. 
    $N$: batch size.}
    \label{fig:network_arch}
\end{figure}

Our baseline network is U-Net~\cite{Ronneberger:2015} with input features $[64, 128, 256, 512]$, and we replace its encoder with a pre-trained low computationally demanding model, MobileNet v2. MobileNet is specifically optimized for efficient execution on low-power devices, making it ideal for mobile and edge applications. The pre-trained MobileNet has weights from ImageNet \cite{Russakovsky:2015}. As Fig.~\ref{fig:network_arch} shows, following Howard et al. \cite{Howard:2018}, we utilize 2D convolutional and bottleneck layers in constructing the MobileNet v2 model. The bottleneck layer, consists of three operations. First, a $1 \times 1$ convolutional layer increases the number of input channels. This is followed by a $3 \times 3$ depthwise separable convolutional layer, which applies a single filter to each input channel. Finally, a $1 \times 1$ convolutional layer is used to decrease the number of channels back to the original dimension.

Given an input image $X \in C \times H \times W$ in HRS, we apply data augmentation on it to get $X_a$. Then we feed $X_a$ into the U-Net network. The feature maps $16 \times \frac{H}{2} \times \frac{W}{2}, 24 \times \frac{W}{4} \times \frac{W}{4}, 32 \times \frac{W}{8} \times \frac{W}{8}, 96 \times \frac{W}{16} \times \frac{W}{16}$ generated from bottleneck layers with channels $C$ in $[16,24,32,96]$ are concatenated to corresponding up-sampling layers in the decoder stack through skip connections. The output from the decoder stack is a prediction matrix $\hat{Z}$ with shape $1 \times H \times W$. After obtaining $\hat{Z}$, we apply $\operatorname{sigmoid}$ to generate the probability matrices. Finally, we calculate the binary cross-entropy (BCE) loss of the probability matrix and ground truth map to optimize the trainable parameters in the network \cite{Ma:2021}. The value of each predicted pixel is binary, either 0 or 1. The optimizer is AdamW with first and second moments of $0.9$ and $0.999$.

\subsubsection{Fine-tuning Strategies}

\begin{figure}[tb]
    \centering
    \includegraphics[width=\textwidth]{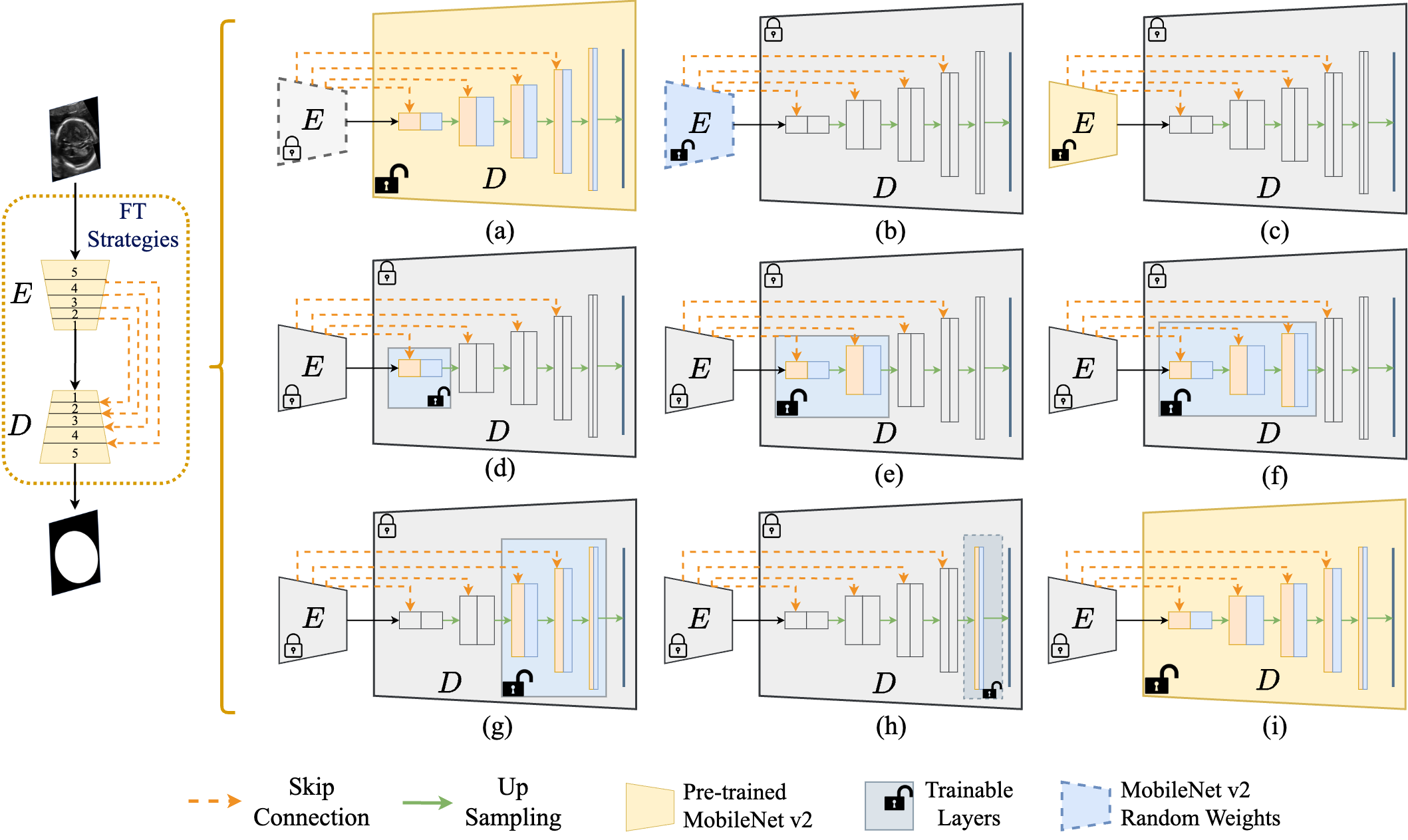}
    \caption{The nine fine-tuning strategies. ($a$) freezes the entire encoder with random weights and trains the decoder; ($b$) trains the encoder with random weights, but freezes the decoder; ($c$) FT\_Encoder strategy trains the entire encoder with pre-trained ImageNet weights; ($d$) trains decoder layer 1; ($e$) trains decoder layers 1,2; ($f$) trains decoder layers 1,2,3; ($g$) trains decoder layers 3,4,5; ($h$) trains decoder layer 5; and ($i$) FT\_Decoder strategy trains decoder layers 1,2,3,4,5. $E$: Encoder; $D$: Decoder; FT: Fine-tuning.}
    \label{fig:ft_methods}
\end{figure}

The fine-tuning strategies in this study include distinct individual layers as well as grouped layers, shown in Fig.~\ref{fig:ft_methods}. Strategies ($a$) and ($b$) utilize MobileNet v2 architecture in the encoder stack with randomly initialized weights. The remaining strategies employ MobileNet v2 architecture with pre-trained weights from ImageNet. The decoder stack consists of 5 layers.

\subsubsection{Statistical Analysis}

To study the model's transferability from HRS to LRS, we design two approaches. The first approach uses a single source in HRS and the second approach includes also samples in LRS. We fine-tune the pre-trained U-Net decoder stack with the HC18 data set using a different amount of samples ($n = 79, 239, 399, 799$). We assess the segmentation performance of the model on both the Spanish and African datasets to evaluate its generalizability in HRS and LRS, respectively. With this, we examine the impact of varying sizes of training data on the performance of the model in HRS and LRS. We utilize the analysis of variance (ANOVA) methodology to quantitatively assess the differences between various fine-tuning strategies and training the U-Net from scratch.

\section{Experiments}

\subsubsection{Dataset} 

\begin{figure}
    \centering
    \includegraphics[width=\textwidth]{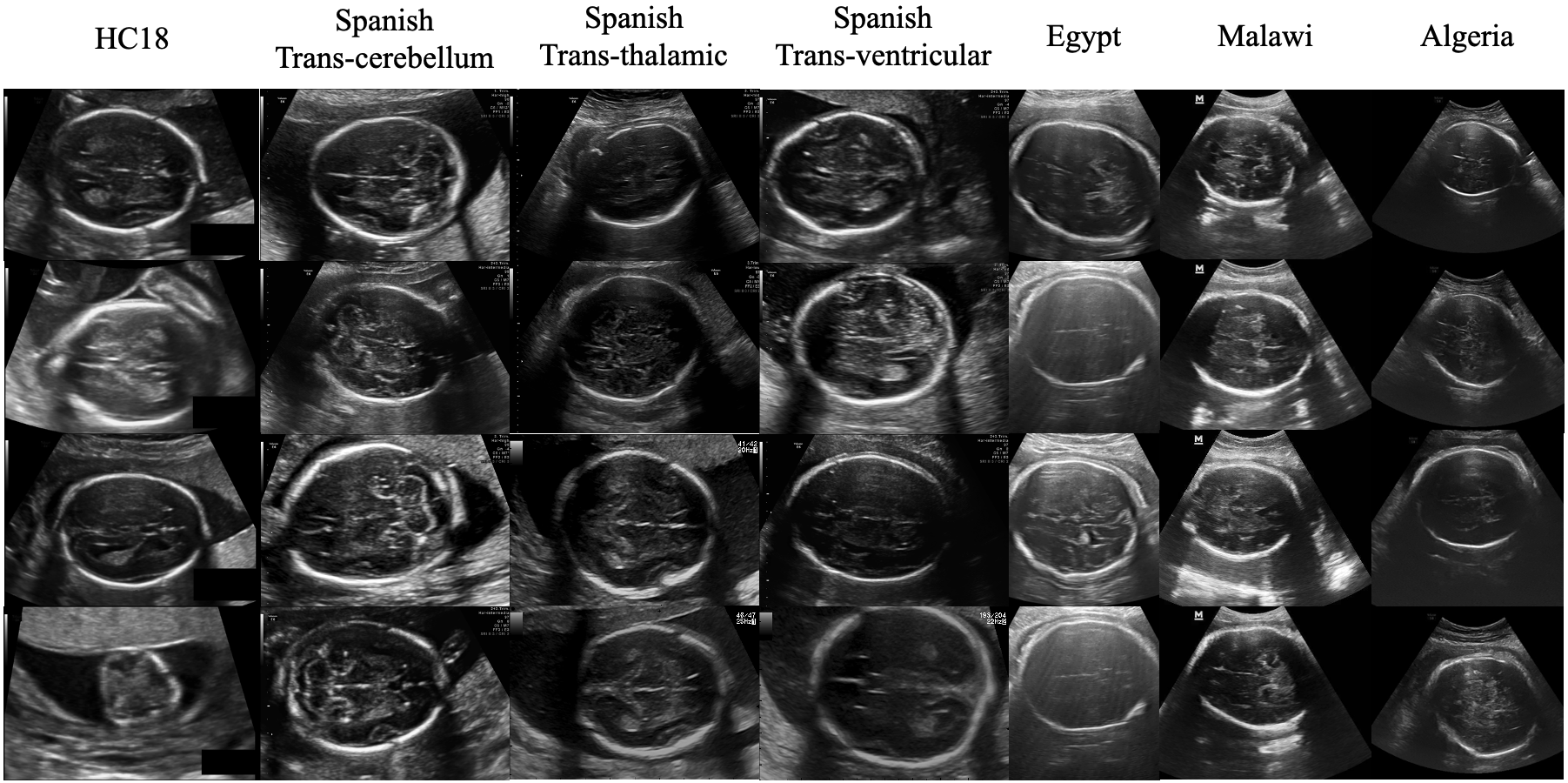}
    \caption{Examples of the maternal-fetal head US images from our multi-centre data set.} \label{fetal_head_samples}
\end{figure}

This study utilizes five public data sets: (1) HC18 from Radboud University Medical Center, Netherlands \cite{Heuvel:2018}. (2) The Spanish data set is acquired from two centres in Spain \cite{Xavier:2020}. (3) The African data sets are from Malawi, Algeria, and Egypt \cite{balcells_data:2023}. Fig.~\ref{fetal_head_samples} illustrates differences (e.g. acoustic shadows, speckle noise, and missing boundaries) between these US images. The input pixels of all images are converted from integer in range $[0, 255]$ to single float precision in $[-1, 1]$ for training and testing purposes. HC18 dataset is published with annotations done by an experienced sonographer and a medical researcher. The Spanish and African cohorts data used in this study are annotated by a medical imaging researcher and reviewed by a clinical expert.

\paragraph{HC18} This dataset includes 2D US data that has been split into 799 images for training purposes and 200 images for validation. All HC18 images are standard planes that are suitable for measuring fetal HC. Each image is of dimensions 800 $\times$ 540 pixels. All images are annotated with biometrics by experienced medical experts. And they are resized to 512 $\times$ 512 pixels for both training and testing purposes. We use the standard data-augmentation techniques: rotation by an angle from $[-25^{\circ}, 25^{\circ}]$, horizontal flipping with 50\% probability ($P(\cdot) = 0.5$), vertical flipping with 10\% probability ($P(\cdot) = 0.1$), and pixel normalization to float precision in $[-1, 1]$ range.

\paragraph{Spanish Data} This dataset includes 1,792 patient records in Spain \cite{Xavier:2020}. All images are acquired on six different machines by operators with similar expertise. These US machines include three Voluson E6, one Voluson S8, one Voluson S10 (GE Medical Systems) and one Aloka. The dataset includes US images for different common brain planes: Trans-thalamic, Trans-cerebellum, and Trans-ventricular. The Trans-thalamic plane includes the frontal horns of the lateral ventricles, cavum septi pellucidi (CSP), the thalami and the hippocampal gyruses \cite{milani:2019}. The Trans-cerebellum plane includes the frontal horns of the lateral ventricles, CSP, thalami, cerebellum and cisterna magna \cite{milani:2019}. The Trans-ventricular plane shows the anterior and posterior portion of the lateral ventricles \cite{milani:2019}. 

\paragraph{African Data} This dataset in LRS is our second testing datasets. It contains images from three African countries \cite{balcells_data:2023}. The Malawi data is acquired at Queen Elizabeth Central Hospital using a Mindary DC-N2 US machine during the 2nd and 3rd trimester \cite{sendra_balcells:2023}. The Egypt data set is obtained at Sayedaty centre using a Voluson P8 (GE), a GE Voluson series ultrasound with less image quality. Only the 2nd-trimester ultrasound scans are acquired \cite{sendra_balcells:2023}. The Algeria data set is acquired during the 2nd and 3rd trimesters at the EPH Kouba and Clinique Des Lilas centres using a lower-quality GE machine \cite{sendra_balcells:2023}. The variability of the fetal head across various pregnancy trimesters poses a challenge in developing a robust DL model.

\subsubsection{Implementation Details}

We build all models in Python with library Segmentation Models~\cite{Iakubovskii:2019}. All training processes are conducted with PyTorch on an NVIDIA Tesla T4 graphics card. Epoch is always set to 20. Batch size is 10 for both training and testing dataset. The Adam optimiser is used in training with a decaying learning rate of \(1e-4\). Training and testing are experimented repeatedly four times for each fine-tuning strategy. In further investigation, we train the baseline U-Net, fine-tuning strategies FT\_Encoder, and FT\_Decoder on 10\%, 30\% and 50\% sub-data sets respectively. To analyse their performance statistically, we apply the analysis of variance (ANOVA) test \cite{fisher:1992}. To evaluate the performance of segmentation, we adopt three typical metrics: Pixel Accuracy ($\operatorname{PA}$), Dice similarity coefficient ($\operatorname{DSC}$), and Mean Intersection over Union ($\operatorname{IoU}$). Mean IoU is defined as the average IoU over all classes \cite{Dice:1945,Jaccard:1912,Tavakoli:2021}.


\section{Results} 

Table \ref{tab:test_diff_ft_methods} shows the competitive segmentation metrics achieved with U-Net baseline, fine-tuning strategy ($a$), ($b$), FT\_Encoder and FT\_Decoder. The fine-tuning strategy FT\_Decoder has contributed to the generation of more accurate predictions on segmentation masks when compared to other methods. This strategy improves PA, DSC, and mIoU by 0.29\%, 0.68\%, and 1.3\% respectively when compared to training the U-Net baseline. Importantly, the size of trainable parameters has been reduced by 85.8\%. The FT\_Decoder strategy demonstrates excellent efficiency when compared to other fine-tuning methods, as it achieves a DSC decrease of only 1\% while utilizing 85\% fewer parameters than Amiri's method \cite{Amiri:2020}, which achieves a DSC of 97\%.

\begin{table}
    \centering
    \caption{Test results from the different fine-tuning strategies. Our proposed fine-tuning strategy achieved the highest average PA, average DSC, and average mIoU. * stands for a MobileNet encoder initialized with random weights.}    
    \begin{adjustbox}{width=\linewidth}
    \begin{tabular}{lccc|ccc}
         \hline
         $E$ Backbone & Params (M) & Strategy & Size (MB) & PA & DSC & mIoU \\
         \hline
         None & 31 & U-Net baseline & 118.5 & 97.33 & 95.56 & 91.56\\
         MobileNet v2* & 4.4 & ($a$) & 59.2 & 91.68 & 86.68 & 76.70\\
         MobileNet v2* & 2.22 & ($b$) & 42.7 & 90.11 & 85.67 & 75.14\\
         MobileNet v2 & 2.22 & FT\_Encoder & 42.7 & 97.54 & 96.05 & 92.42\\
         MobileNet v2 & 4.4 & FT\_Decoder & 59.2 & \textbf{97.77} & \textbf{96.28} & \textbf{92.87}\\
         
         \hline
    \end{tabular}
    \end{adjustbox}
    \label{tab:test_diff_ft_methods}
\end{table}

\begin{figure}[!htb]
    \centering
    \includegraphics[width=\textwidth]{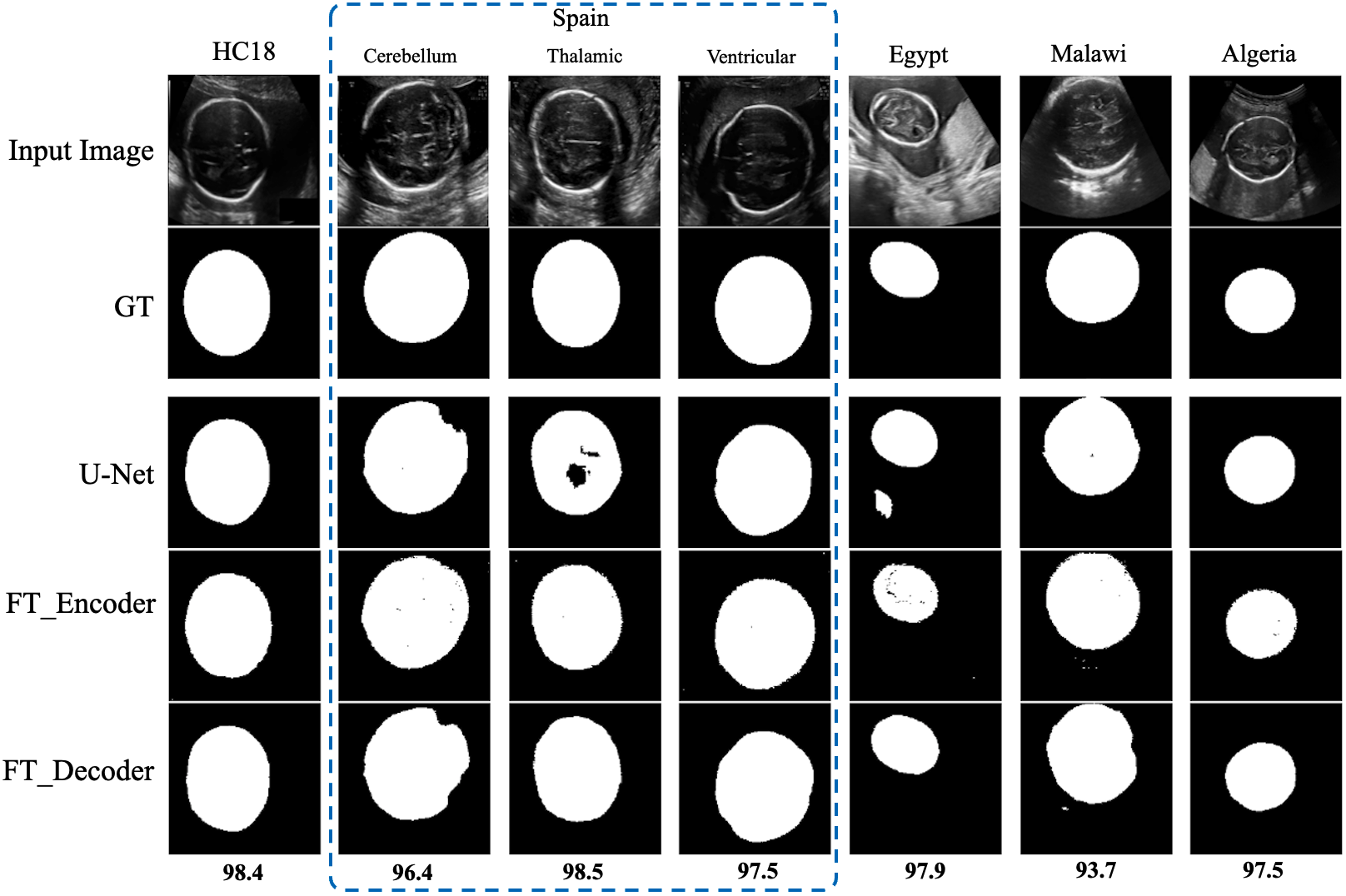}
    \caption{Selected examples of ground truth masks and predicted masks by U-Net baseline and fine-tuning strategies FT\_Encoder, FT\_Decoder respectively. Numbers stand for DSC of each image. 
    }
    \label{fig:test_visual}
\end{figure}

Fig.~\ref{fig:test_visual} illustrates the segmentation masks generated by our fine-tuning strategies. The proposed fine-tuning strategy FT\_Decoder exhibits high DSC in predicting fetal head masks in US images. In the Spanish Trans-cerebellum and Malawi image, the FT\_Decoder method fails to accurately detect the edges of the fetal head with blurred edges. This failure can be attributed to the amplified noise levels present at the edges of the head in these US images, which poses a more formidable challenge to the task of US image segmentation. To the best of our knowledge, statistical metrics for differences across these datasets have not been presented previously yet.

The results of the ANOVA test indicate that training U-Net yields a lower average DSC than strategy FT\_Decoder (mean difference $= -1.70$). And strategy FT\_Encoder exhibits a significant difference compared to strategy FT\_Decoder ($p < 0.05$) with a substantially lower average DSC (mean difference $= -7.87$).

\begin{figure}[!htb]
    \centering
    \includegraphics[width=\textwidth]{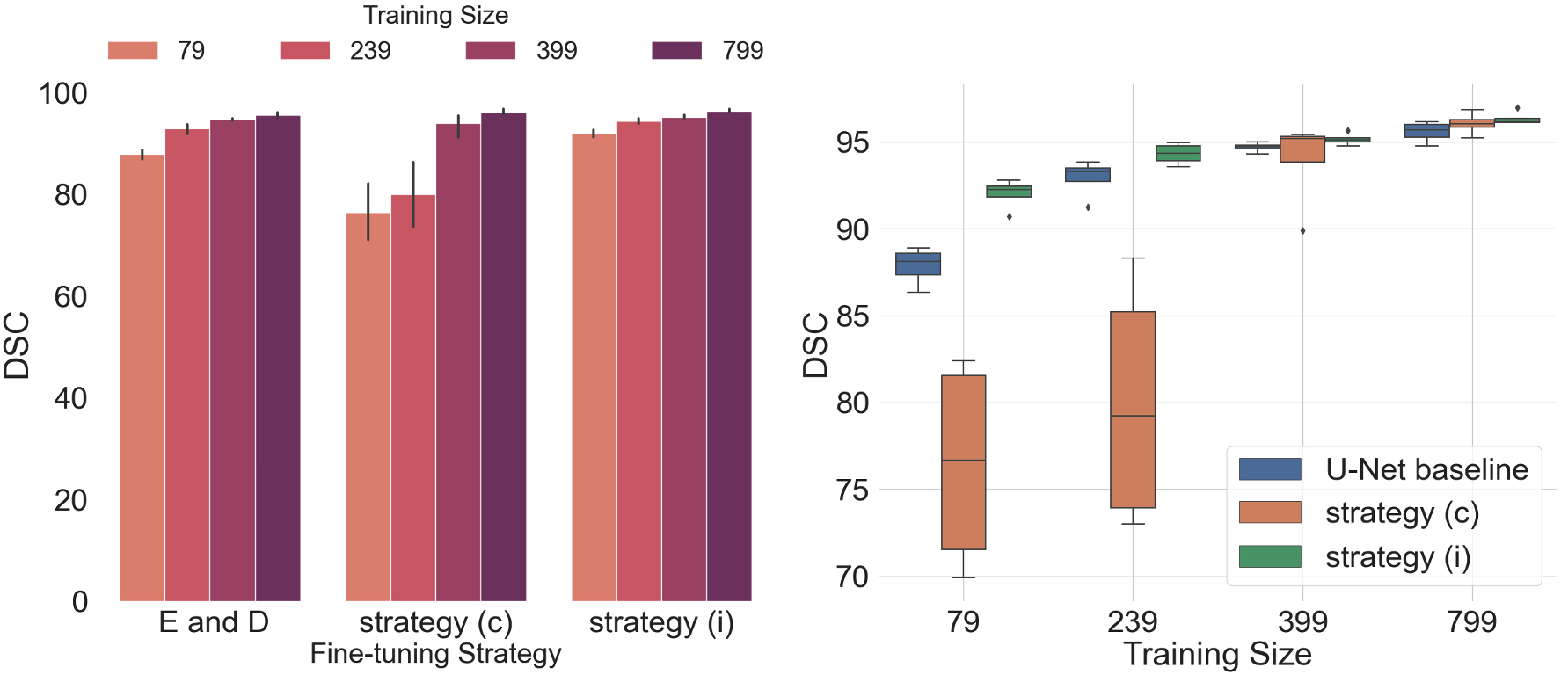}   
    \caption{The results of the U-Net baseline, fine-tuning strategies FT\_Encoder and FT\_Decoder are presented. The \textbf{Left} plot shows the $\operatorname{DSC}$ versus Trainable Layers by Train Size. The \textbf{Right} plot showcases the mean and variance of $\operatorname{DSC}$.}
    \label{fig:performance_plots}
\end{figure}

The segmentation performance of our experiments is numerically visualized in Fig.~\ref{fig:performance_plots}. The left sub-figure demonstrates that the fine-tuning strategy FT\_Decoder is a stable and robust TL method, evidenced by the minimal variance in $\operatorname{DSC}$ across different sizes of training data sets. The $\operatorname{DSC}$ remains relatively consistent, exhibiting little fluctuations. The right boxplot provides a clearer visualization of the stability and robustness of our proposed fine-tuning strategy FT\_Decoder with different training sizes. Furthermore, both plots indicate that among the three strategies we evaluated, fine-tuning strategy FT\_Encoder performs the poorest in fetal head segmentation. 

\subsubsection{Generalization and Transferability}

\begin{table}[!htb]
    \centering
    \caption{Overall performance of Spanish Fetal head US images, African fetal head US images, and HC18 images in metrics of average $\operatorname{PA}$, average DSC, and average $\operatorname{mIoU}$, using U-Net baseline, and strategies FT\_Encoder, FT\_Decoder. The FT\_Decoder shows better performance than other strategies in most of fetal US datasets.}    
    \begin{adjustbox}{width=\linewidth}
    \begin{tabular}{lcccccc}
         \hline
         $E$ Backbone & Strategy & Fetal US Category & Test Size & Avg. PA & Avg. Dice & Avg. mIoU\\
         \hline
         None & U-Net baseline & HC18 & 200 & 97.33 & 95.56 & 91.56\\
         None & U-Net baseline & Cerebellum & 100 & 94.23 & 90.72 & 83.37\\
         None & U-Net baseline & Thalamic & 100 & 94.14 & 90.26 & 82.50\\
         None & U-Net baseline & Ventricular & 100 & 94.89 & 91.89 & 85.14\\
         None & U-Net baseline & Algeria & 25 & 97.48 & 93.10 & 87.12\\
         None & U-Net baseline & Egypt & 25 & 85.62 & 78.51 & 65.11\\
         None & U-Net baseline & Malawi & 25 & 94.77 & 89.70 & 81.42\\
         \hline
         MobileNet & FT\_Encoder & HC18 & 200 & 97.54 & 96.05 & 92.42\\
         MobileNet & FT\_Encoder & Cerebellum & 100 & 96.15 & \textbf{94.42} & \textbf{89.45}\\
         MobileNet & FT\_Encoder & Thalamic & 100 & 96.32 & 94.52 & 89.63\\
         MobileNet & FT\_Encoder & Ventricular & 100 & 96.47 & 94.86 & 90.25\\
         MobileNet & FT\_Encoder & Algeria & 25 & 97.33 & 92.74 & 86.48\\
         MobileNet & FT\_Encoder & Egypt & 25 & \textbf{92.76} & \textbf{90.04} & \textbf{81.98}\\
         MobileNet & FT\_Encoder & Malawi & 25 & 96.13 & 92.53 & 86.11\\
         \hline
         MobileNet & FT\_Decoder & HC18 & 200 & \textbf{97.77} & \textbf{96.28} & \textbf{92.87}\\               
         MobileNet & FT\_Decoder & Cerebellum & 100 & \textbf{96.18} & 94.16 & 89.00\\
         MobileNet & FT\_Decoder & Thalamic & 100 & \textbf{96.89} & \textbf{95.15} & \textbf{90.77}\\
         MobileNet & FT\_Decoder & Ventricular & 100 & \textbf{97.82} & \textbf{96.71} & \textbf{93.63}\\
         MobileNet & FT\_Decoder & Algeria & 25 & \textbf{97.83} & \textbf{94.06} & \textbf{88.79}\\
         MobileNet & FT\_Decoder & Egypt & 25 & 91.30 & 87.50 & 77.89\\
         MobileNet & FT\_Decoder & Malawi & 25 & \textbf{96.33} & \textbf{92.79} & \textbf{86.57}\\
         \hline
    \end{tabular}
    \end{adjustbox}
    \label{tab:more_test}
\end{table}

Table~\ref{tab:more_test} presents the average $\operatorname{PA}$, average $\operatorname{DSC}$, and average $\operatorname{mIoU}$ for fetal head segmentation across three planes (Trans-cerebellum, Trans-thalamic, and Trans-ventricular). The fine-tuning strategy FT\_Decoder achieves the highest DSC for segmenting the head in both the Trans-thalamic and Trans-ventricular planes. It also achieves a DSC of 94.16\% on Trans-cerebellum planes. Training a U-Net from scratch does not demonstrate transferability as effectively as the fine-tuning strategy FT\_Decoder, when implementing transfer learning from HC18 to Spanish US data.

When testing the U-Net baseline, FT\_Encoder and FT\_Decoder on African fetal head US images in LRS, our proposed strategy FT\_Decoder achieves the highest $\operatorname{PA}$, $\operatorname{DSC}$, and $\operatorname{mIoU}$ on images from Algeria and Malawi. It also achieves a DSC of 90.04\% for segmenting fetal heads on US images from Egypt. It is noteworthy that fine-tuning strategy FT\_Decoder has the least decrease in segmentation performance on unseen data in LRS. Table~\ref{tab:more_test} reveals that fine-tuning strategy FT\_Decoder exhibits superior generalization and transferability from HRS to LRS for fetal head segmentation in US images.

\section{Conclusion}

In this work, we present an efficient fine-tuning strategy FT\_Decoder for U-Net to achieve better segmentation performance for fetal head in ultrasound images obtained from low-resource settings. Strategy FT\_Decoder only requires training a decoder stack within a U-Net, achieving a reduction of 85.8\% in the trainable parameters and completing the training in 20 epochs. Our statistical analysis reveals that the FT\_Decoder improves the average DSC by 1.7\% and 7.87\%, respectively, when testing images from HRS and LRS. Our study demonstrates the practicality of fine-tuning a U-Net with limited resources for segmenting fetal head in ultrasound images.

%
%


%
%
%
\bibliographystyle{splncs04}
\bibliography{mybibliography}
%




\end{document}